# Improved Gradual Resistive Switching Range and 1000× On/Off Ratio in HfO$_x$ RRAM Achieved with a Ge$_2$Sb$_2$Te$_5$ Thermal Barrier

R. Islam, S. Qin, S. Deshmukh, Z. Yu, C. Köroğlu, A. I. Khan, K. Schauble, K. C. Saraswat, E. Pop, and H.-S. P. Wong

*Department of Electrical Engineering, Stanford University, Stanford, CA, 94305, USA*

**ABSTRACT:** Gradual switching between multiple resistance levels is desirable for analog in-memory computing using resistive random-access memory (RRAM). However, the filamentary switching of HfO$_x$-based conventional RRAM often yields only two stable memory states instead of gradual switching between multiple resistance states. Here, we demonstrate that a thermal barrier of Ge$_2$Sb$_2$Te$_5$ (GST) between HfO$_x$ and the bottom electrode (TiN) enables wider and weaker filaments, by promoting heat spreading laterally inside the HfO$_x$. Scanning thermal microscopy suggests that HfO$_x$+GST devices have a wider heating region than control devices with only HfO$_x$, indicating the formation of a wider filament. Such wider filaments can have multiple stable conduction paths, resulting in a memory device with more gradual and linear switching. The thermally-enhanced HfO$_x$+GST devices also have higher on/off ratio (>10$^3$) than control devices (<10$^2$), and a median set voltage lower by approximately 1 V (~35%), with a corresponding reduction of the switching power. Our HfO$_x$+GST RRAM shows 2× gradual switching range using fast (~ns) identical pulse trains with amplitude less than 2 V.

Abundant-data computing requires significant data movement to and from off-chip memory, resulting in a "memory-wall bottleneck," where speed and energy efficiency are dominated by the data movement.[1] In order to solve this memory-wall bottleneck, fine grained access between memory and logic is required[2] for which two types of solutions exist: (i) integrating multi-bit digital memory on-chip with high capacity and (ii) in-memory computing, a type of neuromorphic computing where some part of the computation is done inside the memory, reducing data movement between computing and memory.[3,4] In-memory computing requires storage of analog values, which requires resistive memories to switch gradually between different resistive states.

Among possible candidates for in-memory computing, resistive random-access memory (RRAM) is one of the emerging non-volatile memory technologies that is highly scalable, back-end-of-line (BEOL) compatible, and capable of low energy switching.[5] Filamentary RRAM is a metal/oxide/metal device that operates by forming single or multiple filaments composed of oxygen vacancies created by a soft breakdown in the oxide due to the applied electric field.[6] One of the challenges for filamentary RRAM to switch gradually across a large range of conductance values is the abrupt set process.[7] The filament formation and the subsequent set and reset cycles are due to O$^{2-}$ ion movement to and from the filament to the top electrode (TE), which serves as the oxygen reservoir. In $HfO_x$ RRAM, the $O^{2-}$ ion movement is mostly driven by the electric field (E-field), due to its relatively low hopping activation energy (0.7 eV).[8] The E-field driven ion movement causes a soft oxide breakdown which initiates rapid positive feedback of current and local self-heating, making the set process abrupt.[8]

However, it has been demonstrated that $O^{2-}$ diffusion is thermally controlled, where both the lateral temperature gradient away from the filament and the high temperature of the filament increase the lateral diffusion of $O^{2-}$.[8,9] Padovani *et al.* demonstrated by kinetic Monte-Carlo modeling that high temperature formation causes a wider filament.[8] Trap-assisted tunneling transport between the oxygen vacancies in a wider filament result in multiple stable conduction paths through the filament resulting in a gradual and linear change in resistances.[10] High temperature operation of the RRAM device as performed by Jiang *et al.*[9] requires a separate micro-thermal stage that is not scalable in practice. Also, this approach cannot directly probe temperature gradients at nanoscale resolution due to fabrication constraints. In addition, probing the surrounding oxide may not be sufficient to get a complete thermal picture of the filament, since majority of the heat is dissipated through the electrodes. Note that the filament typically ruptures and reforms at the top electrode interface. Adding a thermal barrier material with low thermal conductivity and good electrical conductivity between the switching oxide and the bottom electrode (BE) could raise the temperature in the switching layer without reducing the E-field within the filament, thereby increasing lateral diffusion of $O^{2-}$ ions to form a wider filament.

Wu *et al.*[11] reported thermal enhancement using a conductive $TaO_x$ layer between $HfO_x$ and the BE, where the gradual switching range from the high to low resistance state (HRS to LRS) is 3× and the switching is highly non-linear. However, no experimental visualization of the wider filament caused by the $TaO_x$ thermal barrier has been reported. Compared to the thermal conductivity of $HfO_x$ (~0.5 to 1.0 $Wm^{-1}K^{-1}$),[12] the $TaO_x$ thermal conductivity is not low (~10 $Wm^{-1}K^{-1}$).[11] Therefore, the origin of gradual resistive switching from $TaO_x$ insertion could be due to the addition of thermal interfaces: electrode/$TaO_x$ and $TaO_x$/$HfO_x$, and the low oxygen vacancy mobility of $TaO_x$ compared to $HfO_x$[13] that could also result in effective width modulation of the conductive filament.[7]

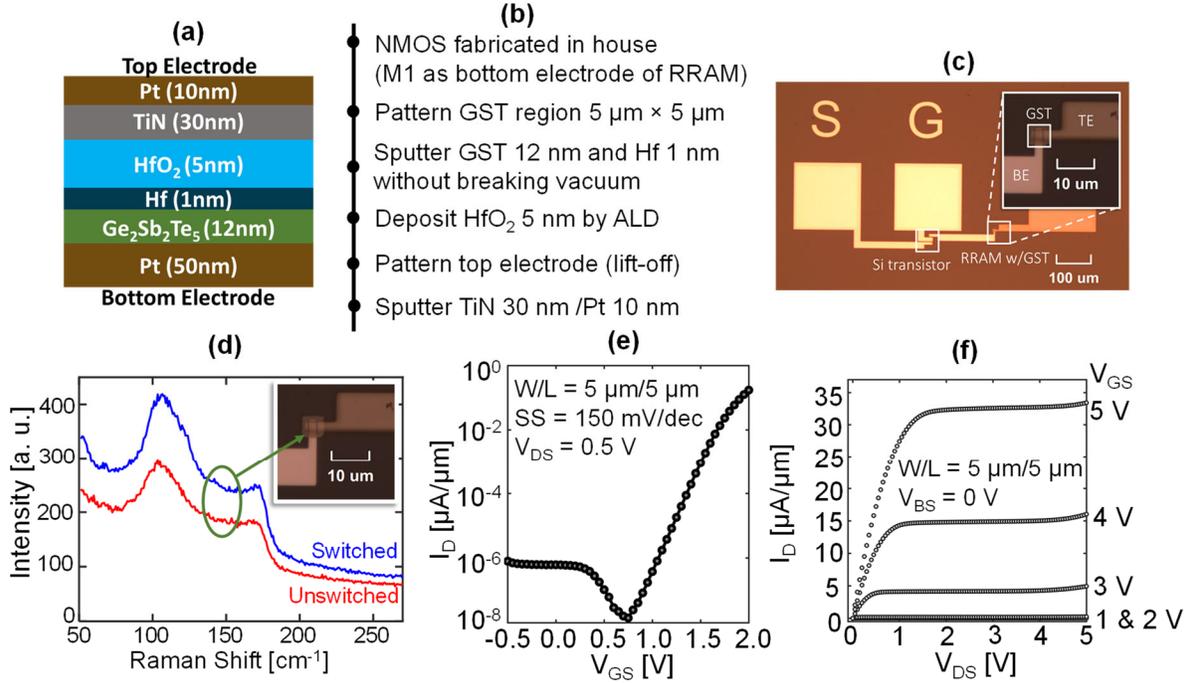

**FIG. 1.** (a) Device cross-section, (b) fabrication flow, (c) Optical image of 1T1R structure. S and G correspond to the source and gate of the access transistor. Note that the GST region extends slightly beyond the cross-bar area to cover the bottom electrode sidewall. (d) Raman spectra of the GST adjacent to the crossbar, indicating crystalline GST both before and after switching. We observe a slight change in the spectra after switching. Inset displays an optical image with the arrow pointing to the spot where Raman measurements were done. (e) Measured $I_{DS}$ vs. $V_{GS}$ showing subthreshold slope (SS) = 150 mV/dec and (f) $I_D$ vs. $V_{DS}$ characteristics of the transistors in 1T1R structure.

In this work, we report experimental visualization of a conductive filament and correlate its morphology with the switching characteristics including on/off ratio, set voltage, and gradual switching behavior. Our experiments suggest that the filament is wider in $HfO_x$ RRAM with a $Ge_2Sb_2Te_5$ (GST) thermal barrier placed between $HfO_x$ and the BE. We choose GST here because such chalcogenide glasses have lower thermal conductivity (~0.45 $Wm^{-1}K^{-1}$ in the fcc phase)[14] and higher electrical conductivity than transition metal oxides (like $HfO_x$, $TaO_x$). We have observed that for similar input power, the $HfO_x$+GST devices also show 1.5-2× higher temperature difference at the filament hot spot with respect to the ambient. This reduces the set voltage and results in a linear and gradual resistive switching due to thermal enhancement in this RRAM device.

Our RRAM is fabricated in series with an NMOS Si transistor to form a 1-transistor 1-resistor (1T1R) test structure. Figures 1(a,b) show a schematic of the RRAM cross-section and the fabrication flow, respectively, and Fig. 1(c) shows the top-side optical image of the completed 1T1R. The transistor has gate length $L$ = 5 μm and width $W$ = 5 μm. After the transistor is fabricated, the drain contact is extended to form the BE (50 nm Pt) of the RRAM, followed by a layer of sputtered GST (12 nm) on the BE. The GST region is 5 μm × 5 μm patterned using a lift-off process to fully cover the BE (1 μm × 1 μm) sidewalls. A very thin capping layer of Hf (1 nm) is sputtered *in situ* to prevent the oxidation of the GST, and the thin Hf film oxidizes to $HfO_x$ on vacuum break. The $HfO_x$ (5 nm) switching layer is then immediately deposited by atomic layer deposition (Cambridge Nanotech Savannah S200) at 200 °C, using TDMA-Hf as Hf precursor and water as the oxygen source. Finally, the 500 nm × 500 nm TE is sputtered and patterned as TiN (30 nm) capped by Pt (10 nm) to make a crossbar structure. Due to the processing temperature of $HfO_x$, the as-deposited amorphous GST crystallizes into the cubic (fcc) phase. In Fig. 1(d) we observe that after switching, the Raman spectra show a slightly asymmetric peak at ~100 $cm^{-1}$ which indicates some mixed phase other than fcc.[16]

The devices undergo RRAM forming with a linear bipolar DC current-voltage (*I-V*) sweep up to 6 V while keeping the BE at 0 V. During the forming process the compliance current is controlled by the transistor gate voltage. The $I_D$ vs. $V_{GS}$ and the $I_D$ vs. $V_{DS}$ characteristics of the NMOS transistor are shown in Figs. 1(e,f) respectively. We measure the apparent changes in the width of the filamentary region with multiple steady-state voltage biases using scanning thermal microscopy (SThM),[16] a scanning probe technique which enables temperature measurement of the RRAM top surface with sub-100 nm spatial resolution.[15,16]

Figure 2(a) shows the SThM tip voltage ($V_{SThM}$, proportional to the surface temperature rise) across the top surface, for devices with only $HfO_x$ and those with $HfO_x$+GST, at several bias conditions. These devices were formed and cycled 5 times to the LRS before measurement. During measurement, a low magnitude, steady-state bias is applied so that the filament conducts a current small enough to not disturb it, but large enough to induce some Joule self-heating which is detected by the SThM on the top surface. Figures 2(a) (i and ii) show the SThM tip voltage map for zero current flowing through the filament. This figure establishes the baseline measurement that represents the topography of the device surface. At a smaller current level, driving 20-30 μW power through the filament, we start to observe the increase in $V_{SThM}$ [Figs. 2(a) (iii) and (iv)] indicating Joule heating. When the power reaches 55-70 μW [Figs. 2(a) (v) and (vi)], a hot spot is seen having the highest $V_{SThM}$. This represents the possible location of the filament. Figure 2(b) shows the estimated Δ*T* at the top of the device, after calibration of $V_{SThM}$ from known temperatures.[16] (This temperature corresponds to the highest $V_{SThM}$ point in the 2D map.)

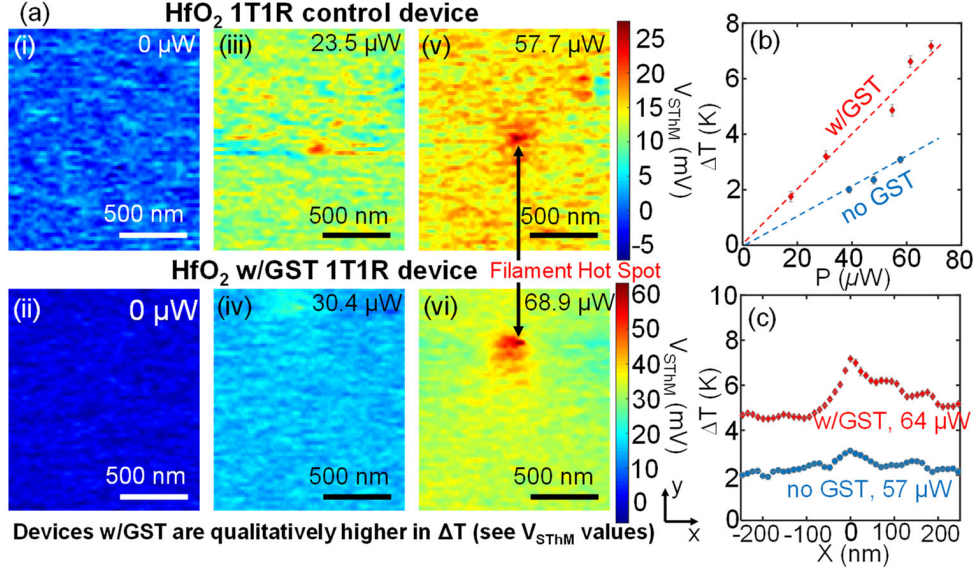

**FIG. 2.** (a) Scanning thermal microscopy (SThM) imaging of the 1T1R devices after filament formation. Top row (i, iii, v) shows a control device with only HfO$_x$, bottom row (ii, iv, vi) shows a HfO$_x$+GST device. The current is kept constant during scanning to ensure the filament conducts, making it visible through the top-side hot spot. Devices with GST have qualitatively higher temperature rise ($\Delta T$), which is directly proportional to the measured $V_{SThM}$ values. (b) Peak temperature at the hot spot relative to the ambient temperature as a function of input power during the scanning. Note that the rate of heating in the HfO$_x$+GST device is 1.5× - 2× that of HfO$_x$ device, (c) Measured temperature profile at the hot spot for a certain power, showing the broadening of the hot spot. This demonstrates the widening of the filament.

We observe that the HfO$_x$+GST device has higher peak hot spot temperature rise ($\Delta T$, which is proportional to $V_{SThM}$), indicating better heat trapping due to the lower thermal conductivity of the GST thermal barrier. The temperature rise at the top surface for the same electrical power is as much as twice in our RRAM device with GST, compared to control devices without, at similar applied electrical power. Figure 2(c) shows the line profile of $\Delta T$ as a function of the *x*-axis [in Fig. 2(a)] through the hot spot. The broader hot spot in HfO$_x$+GST RRAM is due to the higher thermal resistance of the HfO$_x$+GST stack that results in a higher temperature rise for the same applied power compared to the GST-only device [Fig. 2(b)]. This suggests that GST with its lower thermal conductivity acts as a thermal barrier which prevents heat loss from the switching layer (HfO$_x$) to the BE. As a result, the HfO$_x$+GST device has higher temperature forming, causing a wider filament. In the HfO$_x$-only device, forming the filament is dominated mostly by the E-field that results in a narrower, more compact filament formation.

One important consequence of a wider filament is the transport mechanism in both set and reset is dominated by trap-assisted tunneling (TAT) between oxygen vacancies spaced farther apart than in a single strong filament [Figs. 3(a) and 3(b)]. TAT is highly sensitive to the average tunneling distance and hence to the microscopic configuration of the filament after each cycle. Figure 4(a) shows the comparison of LRS and HRS distributions for both HfO$_x$-only and HfO$_x$+GST devices for 50 switching cycles. TAT dominated transport in the wider filament in HfO$_x$+GST device show a broader resistance distribution (more prominent in HRS) and a higher on/off ratio. Note that the on/off ratio increases with the compliance current for HfO$_x$+GST device whereas, it remains very similar for HfO$_x$-only device. The energy barrier for electrons to hop from one trap state to another depends on the E-field.[17] Therefore, switching in conventional HfO$_x$-only RRAM is determined by E-field dependent direct tunneling during set and by Joule heating during reset, where the Joule heating enhances oxygen transport. On the other hand, our RRAM with the GST thermal barrier switches by a combination of tunneling transport mechanisms during both set and reset. This is due to the higher temperature forming that causes increased defect density in the oxide surrounding the filament. Heat trapping also helps move oxygen ions relatively easily during set in response to the applied E-field. In other words, higher temperatures make the oxygen ion more responsive to E-field.

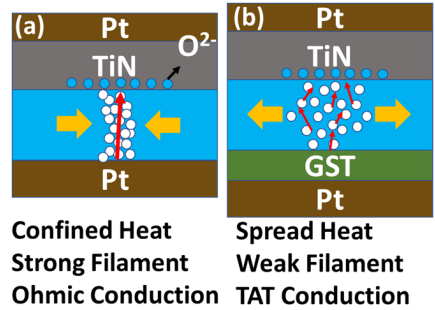

**FIG. 3.** Schematic of the filament in (a) HfO$_x$-only (b) HfO$_x$+GST device. A wider filament is governed by trap-assisted tunneling (TAT) conduction.

Despite the insertion of a relatively thicker GST layer with respect to HfO$_x$, the set voltage ($V_{set}$) does not increase because the HfO$_x$ is more resistive than the GST when the device is in HRS. A majority of the $V_{set}$ voltage drop is across the oxide rather than the GST, because GST is more electrically conductive. In fact, Fig. 4(b) shows that the $V_{set}$ of the HfO$_x$+GST device is lower than that of the HfO$_x$ device (the distribution is over 50 cycles of switching) for same reset voltage ($V_{rst}$). This also suggests that adding GST makes it easier to form a filament by heat trapping inside the HfO$_x$, in agreement with the findings of Jiang *et al.*[9] that forming voltage is lowered at higher temperatures. The reduction in $V_{set}$ is more significant for higher compliance current ($I_{cc}$ = 100 µA), which indicates that more heating in the filament increases the mobility of the O$^{2-}$ ions, requiring less E-field for set.

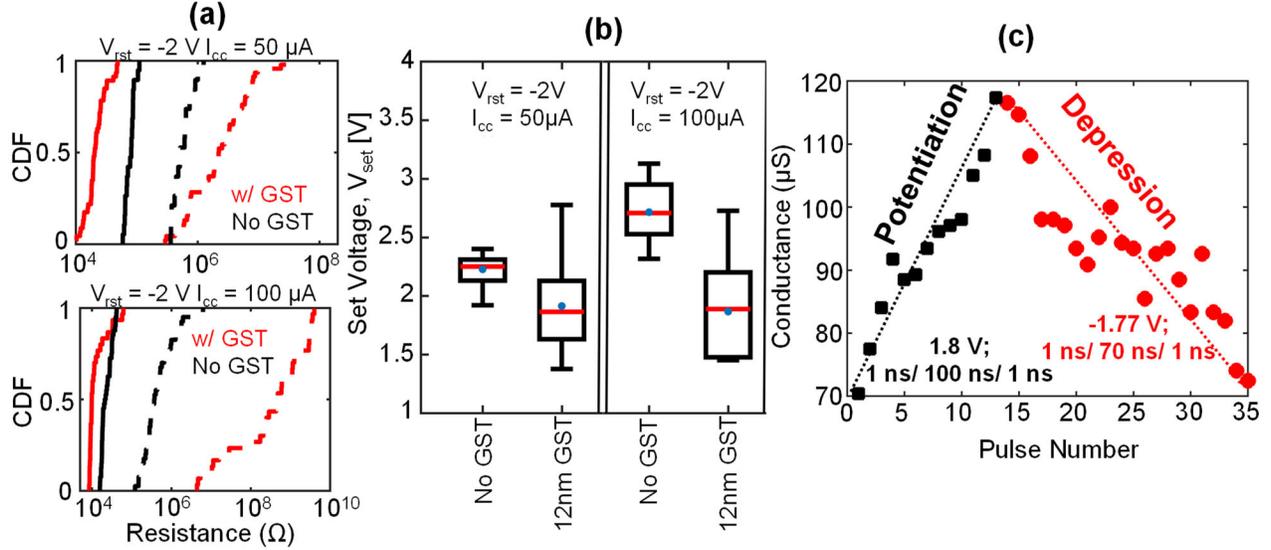

**FIG. 4.** (a) Cumulative distribution function (CDF) of LRS and HRS states measured at a read voltage of 0.1 V after each DC switching cycle of the RRAM, calculated over 50 switching cycles, comparing a $HfO_x$+GST device (red lines) with a $HfO_x$-only device (black lines) for different compliance currents, $I_{cc}$. Solid lines represent LRS and dashed lines represents HRS. Note the on/off ratio increases significantly for the $HfO_x$+GST device. (b) Distribution of set voltages ($V_{set}$) during each DC switching cycle extracted over 50 switching cycles of the RRAM, comparing a $HfO_x$+GST device with a $HfO_x$-only device for different compliance currents ($I_{cc}$). (c) Conductance vs. pulse number for a sequence of identical pulses applied to the $HfO_x$+GST 1T1R device. The pulse shapes are marked on the figure.

We demonstrate gradual switching in $HfO_x$+GST RRAM by applying a pulse train with the same amplitude and width, and measuring the resistance after each pulse. The conductance as a function of the number of pulses is shown in Fig. 4(c). Both potentiation (low to high conductance) and depression (high to low conductance) show gradual and linear switching with a ~2× dynamic range. For potentiation (depression), the write pulse amplitude and width are +1.8 V (-1.77 V) and 100 ns (70 ns), respectively. In both cycles, a 1 ns rise/fall time is used. The gradual switching response is linear with respect to the number of pulses applied. This demonstrates that heat trapping in $HfO_x$ due to the GST thermal barrier layer causes gradual set/reset.

In conclusion, we have demonstrated experimental observation of filament formation in RRAM, and observed that inserting a GST thermal barrier causes a wider filament to form in the $HfO_x$. Such thermally-enhanced RRAMs show higher on/off ratio and linear gradual resistive switching. Gradual resistance switching is promoted by the formation of wider instead of narrow filaments (in control devices without a GST barrier, where the oxygen vacancies are more compact). Such change in the filament morphology can increase the contribution of trap-assisted tunneling, which reduces the bistable nature of filamentary switching. Lateral heat spreading inside the $HfO_x$ switching material also improves the mobility of $O^{2-}$ ions, resulting in a decrease of the set voltage, which can potentially improve the energy-efficiency of switching. Such thermally-enhanced RRAMs can be a potential candidate as synaptic devices for in-memory computing.

This work was supported in part by ASCENT (one of six centers in JUMP, a Semiconductor Research Corporation (SRC) program sponsored by DARPA), and member companies of the Stanford SystemX Alliance and the Stanford Non-Volatile Memory Technology Research Initiative (NMTRI). Part of this work was performed at the Stanford Nanofabrication Facility (SNF) and Stanford Nano Shared Facilities (SNSF), supported by the National Science Foundation under awards ECCS-1542152 and ECCS-2026822.

**AUTHOR DECLARATIONS**

**Conflict of Interest:** The authors have no conflicts to disclose.

**DATA AVAILABILITY**

The data that support the findings of this study are available from the corresponding author upon reasonable request.